\documentstyle[11pt,newpasp,twoside,epsf]{article}
\def\edcomment#1{\iffalse\marginpar{\raggedright\sl#1\/}\else\relax\fi}
\reversemarginpar

\begin{document}
\title{A New Component of the Galaxy as the Origin of the 
LMC Microlensing Events}
 \author{Evalyn Gates}
\affil{Adler Planetarium \& Astronomy Museum, Chicago, IL 60605}
\affil{Department of Astronomy \& Astrophysics, University of Chicago, Chicago,
IL 60637}
\author{Geza Gyuk}
\affil{Physics Department, University of California, San Diego, CA 92093}

\begin{abstract}
We suggest a new component of the Milky Way that can account for both
the optical depth and the event durations implied by microlensing searches
targeting the Large Magellanic Cloud.  This component, which 
represents less than
$4\%$ of the total dark matter halo mass, consists of mainly
old white dwarf stars in a distribution that extends
beyond the thin and thick disks, but lies well within the dark matter halo. 
It is consistent with 
recent evidence for a significant population of faint white dwarfs detected
in a proper motion study of the Hubble Deep Field
that cannot be accounted for by stars in the disk or spheroid.
Further, it evades all of the current observational constraints that 
restrict a halo population of white dwarfs.
\end{abstract}

\section{Introduction}

The past few years have yielded exciting new data from
observational teams searching for evidence of MACHOs  
(dark compact objects) in the halo 
of our galaxy, data which have raised many new questions. The microlensing
optical depth toward the Large Magellanic Cloud (LMC) 
obtained by the MACHO collaboration,
$\tau_{\rm LMC} = (1.2^{+0.4}_{-0.3})\times 10^{-7}$ (Cook et al. 2000),
is consistent with a significant fraction of the galactic halo ($\sim 20\%$)
in the form of MACHOs.  The duration of these events also
indicates that, under the assumption of a spherical isothermal halo 
distribution for the MACHOs, the average MACHO mass is $ \sim 0.5 M_\odot$ 
(with large statistical uncertainties).                

Such masses suggest several candidates for the lenses including faint 
halo stars, white dwarfs, and black holes.  
Direct searches for faint
halo stars have placed severe limits on their
contribution to halo, requiring it to be less than about $3\%$ 
(Flynn, Gould, \& Bahcall 1996; Graff \& Freese 1996).
Primordial black hole candidates require a fine tuning of the initial
density perturbations, and details of QCD phase transition
black hole formation remain to be worked out in order to assess
their viability and mass function (Jedamzik \& Niemeyer 1999).  
Thus in the standard
halo model interpretation, white dwarfs appear to be the remaining
strong candidate for the lenses. 

However, other possible interpretations of the MACHO results have been
proposed in order to avoid the difficulties associated with a large halo
population of white dwarfs.  The mass estimate for the lenses depends upon
the assumed phase space distribution of MACHOs. There are large
uncertainties in the halo model parameters, including the distribution and
velocity structure of the dark matter, and attempts have been made to
exploit these uncertainties in order to obtain mass estimates from the
current data that are consistent with lenses in the substellar regime
(e.g. brown dwarfs). Previous work by the authors and others have examined
a wide range of halo models, including flattened halos,
halos with a bulk rotational component to the velocity structure
(Gyuk \& Gates 1998) and halos with anisotropic velocity dispersions
(Gyuk, Evans, \& Gates 1998).  These analyses have shown that for any 
reasonable
(smoothly varying) phase space distribution of the lenses, the implied
lens mass is still much larger than the hydrogen burning limit, and thus
one cannot appeal to modeling uncertainties in order to invoke brown
dwarfs as candidates for the MACHOs. 

Other work has explored the possibility that the lenses are not in the
halo of the Milky Way.  LMC self-lensing was originally suggested by Sahu
(1994), but recent work by Gyuk, Dalal \& Griest (1999) has argued that
this is unlikely to be able to account for the current data. 
More recently, an unvirialized component
of the LMC has been proposed in order to produce sufficient LMC self-lensing
(Graff 2000). Zaritsky \& Lin (1997) and Zhao (1998) have suggested
an intervening population of stars toward the LMC (tidal debris or a dwarf
galaxy) could be responsible for the microlensing events. Again this
suggestion has been subject to much debate, and a recent paper by Gould
(1999) argues strongly against such a scenario.  Galactic models in which
dark extensions of known populations, such as a heavy spheroid or thick
disk, could be the source of the lenses were explored by Gates, Gyuk,
Holder \& Turner (1998). 

However, recent results from a proper motion study of the Hubble Deep
Field (HDF) (Ibata et al. 1999 ), and from a comparison of the north and south
HDF images (Mendez \& Minniti 1999) have added a new piece to the puzzle. These
studies provide further evidence that there may be a previously undetected
population of old white dwarfs in the galaxy. Using new models for white dwarf
cooling (Hansen 1999), the sources
detected by Ibata et al. are consistent with 2-5  old ($> 12 Gyr$), $0.5
M_{\odot}$ white dwarfs with kinematics roughly consistent with that 
expected for halo objects. While earlier searches for the local counterparts 
did not find any such objects (Flynn et al. 2000), recent surveys have 
turned up several candidates (Hodgkin et al. 2000; Ibata
et al. 2000). Overall, these data strengthen the
interpretation of the MACHO lenses as white dwarfs in our own galaxy, 
and thus make alternative scenarios, such as LMC self-lensing, less appealing.
 
While these new data are still somewhat preliminary, they do raise the
intriguing possibility that there is a previously undetected population of
white dwarf stars that are not part of the disk or (known) spheroid.
Along with the MACHO data and the inability of modeling
to significantly change the mass estimates, these new results, seem to 
be relentlessly
pointing to white dwarfs as the lenses.  So, is the halo of our galaxy
filled with white dwarfs? A standard halo interpretation of these data
would say yes -- a significant fraction of the galactic halo must be in
white dwarfs.  However, such a scenario faces serious challenges from many
directions.

\section{White Dwarfs in the Halo?}

When considering the possibility that a large fraction (or all) of the
galactic halo might be in the form of white dwarfs, it is extremely
important to recall the evidence for galactic dark matter, including
estimates of the total mass of the Milky Way.  A recent analysis of
satellite radial and proper motions by Wilkinson and Evans (1999) found a
total mass of the Galaxy $M_{TOT} \sim 2\times 10^{12} M_\odot $, in good
agreement with other recent estimates (Kochanek 1996; Zaritsky 1998).
Wilkinson and Evans also find that the halo extends to at least $100
\rm{kpc}$, and possibly much further to 150 or 200 kpc.  Thus the total
mass in a white dwarf population that comprises a significant fraction of
the halo would be of order $10^{12} M_\odot$, a number which already
severely strains the baryon budget of the Universe.

Models which propose such a population must also account for the mass in
the progenitor population of stars and in the metal enriched gas produced
during the formation of the white dwarfs.  Combined with the above mass
estimate for the galactic halo, such considerations provide serious
challenges for these models.  For example, consider a white dwarf halo
which is comprised of at least $50\%$ white dwarfs.  The total mass in
white dwarfs today is thus of order $10^{12}$.  The efficiency $\epsilon
(m)$ for producing a white dwarf from a progenitor star of mass $m$ is
likely to be 0.25 or smaller, depending on the progenitor mass (with an
upper limit of $\epsilon = 0.5$ for progenitor stars of $1 M_{\odot}$)
(Adams \& Laughlin 1996).  Thus, for a white dwarf halo mass of $M_{wd}$, 
we expect
a mass in the progenitor population $M_{Pstars} \geq 4M_{wd}$ and a mass
in processed, metal rich gas $M_{zgas} \geq 3M_{wd }$.  A halo of mass
$M_{TOT} = 2 \times 10^{12}$, half of which is in white dwarfs, requires a
progenitor mass of $M_{Pstars} \geq 4 \times 10^{12}$.  This in turn
requires an extremely efficient early burst of star formation, through
which essentially all of the baryons in the Universe are processed.

From a cosmological point of view, we can consider the contribution of
the white dwarfs and the progenitor population to the matter density of
the Universe.  The Milky Way has a mass to light ratio $M/L \sim 100$ or
greater (Zaritsky 1998).  If we assume that this is a typical value for
all galaxies, then galaxies contribute $\Omega_{g} \ga 100/1200h = 0.08
h^{-1}$.  Comparing this with $\Omega_{b}h^2 = 0.019 \pm 0.0024$
(95$\%$cl, Burles et al. 1999), we find that a $50\%$ white dwarf 
halo exceeds the
baryon budget (${\Omega_{MACHOs}\over \Omega_b} \sim 2 h$) even before
considering the effects of processing most of the baryons through an early
star phase. A $20\%$ white dwarf halo is also difficult to reconcile with
the above estimate of $\Omega_{b}$, since the contributions of the
progenitor stars will exceed $\Omega_b$.

Many authors have explored the implications of a halo filled with white
dwarfs.  These analyses, combined with the estimates of the total mass in
the halo, make the possibility of a white dwarf halo even less
tenable. 

First, the initial mass function (IMF) of the progenitor stars must be
markedly different than the disk IMF.  Limits on
the IMF arise from both low and high mass stars.  Low mass stars ($< 1
M_{\odot}$) would still be burning hydrogen today and should be visible.
High mass stars ($> 8 M_{\odot}$) would have evolved into Type II
supernovae, ejecting heavy metals back into the interstellar medium
\footnote{However, Venkatesan, Olinto \& Truran 1999 have argued that the
bounds on high mass stars are significantly relaxed for progenitor stars
with very low ($10^{-4} Z_{\odot}$) or zero metallicity, 
which might relax this constraint somewhat.}.  From limits
on red dwarfs in the halo and the galactic metallicity,
Adams \& Laughlin (1996) and Chabrier et al
(1996) find that the IMF must be sharply peaked about a
progenitor star mass of $m \sim 2 M_{\odot}$.  Adams \& Laughlin conclude
that even with the above IMF, the white dwarf contribution to the halo is
limited to less than $25\%$ (with $50\%$ being an extreme upper limit).

Next, the metal enriched gas produced when these stars become white dwarfs
will pollute the remaining unprocessed gas, leading to high metallicities
predicted for the Galactic disk and the interstellar medium (into which
much of this gas must be blown out since the total mass in processed gas
is much larger than the mass of the disk)(Fields, Matthews, \& Schramm 1997).
Gibson \& Mould
(1997) have estimated that the expected amount of C, N and O produced by
the above IMFs, where the white dwarf mass is about $4\times 10^{11}M_{\odot}$
(for a $20\%$ MACHO halo), 
would be difficult to reconcile with that in pop II white dwarfs.
  
White dwarfs in the halo would also produce heavy metals via Type Ia
supernovae.  Canal, Isern and Ruiz-Lapuente (1997) use this to limit the
halo fraction in white dwarfs to less than $5-10\%$ (or a total mass in
white dwarfs of $5-10 \times 10^{10} M_{\odot}$).  In addition, deep
galaxy counts limit the fraction of the halo in white dwarfs to less than
about $10\%$, since the
brightly burning progenitor stars would be visible (Charlot \& Silk 1995).
Finally, it is worth mentioning that an all white dwarf halo would rule
out the existence of other dark matter in the Universe (for example cold
dark matter) (Gates \& Turner 1994), opening the door to a host of problems 
with large scale structure formation.

\section{A New Component of the Galaxy}

Given the evidence for a previously undetected population of white dwarfs
and the severe constraints on a halo population consistent with this
evidence we propose a new component of the Galaxy.  Such a component was
first considered by the authors (Gyuk \& Gates 1999) in the context of
attempting to lower the mass estimates for the MACHO lenses, and
in Gates et al. (1998) in considering dark extensions to know components.

This new component is essentially a very thick (scale height $>2$
kpc) population of (mostly) old white dwarf stars.  It is distinct from
known galactic populations, both in distribution and age.  This ``extended
protodisk'' extends beyond the thin and thick disk populations, but lies
well within the halo.  While the details of the distribution cannot be
determined without significantly more data, the general features of this
proposed model can be illustrated with the following example:

Consider an exponential disk with a volume density given by

\begin{equation}
\rho(r,z) = {\Sigma_0\over 2h_z}exp((r_0 - r)/r_d)sech^2(z/h_z)
\end{equation}
where $r_d = 4.0\rm{kpc}$ is the scale length and $h_z = 2.5\rm{kpc}$ is
the scale height. We assume standard values for the position and circular
velocity of the Sun, $r_0 = 8.0\rm{kpc}$ and $v_c = 220 {\rm km/s}$.

We also assume a velocity structure, which includes a rotational component
$\tilde{v}_\phi \sim 130-170 \rm {km/s}$, of the form
\begin{equation}
f=\frac{\rho(r,\phi,z)}{m}
\frac{1}{\sqrt{(2\pi)^3}\sigma_r\sigma_\phi\sigma_z}
e^{-\left[\frac{v_r^2}{2\sigma_r^2}+\frac{(v_\phi-\tilde{v}_\phi)^2}{2\sigma_\phi^2}+\frac{v_z^2}{2\sigma_z^2}\right]}
\end{equation}

where the velocity ellipsoid varies as 

\begin{equation}
\sigma_z^2 = 2 \pi G \rho_0 h_z^2 = \pi G \Sigma_0 h_z.
\end{equation}

\begin{eqnarray*}
\sigma_r^2 & \approx & 2 \sigma_z^2 \\
\sigma_\phi^2 & \approx & \sigma_z^2.
\end{eqnarray*}
For details on varying these model parameters as well as different
parameterizations of the generic extended protodisk model see
Gyuk \& Gates (1998, 1999). 

We can also consider a 
spheroid-like distribution for this component.
Dynamical estimates for the mass of the spheroid are considerably larger
than the luminous mass, although recent studies of the mass function of
the spheroid indicate that the known spheroid population is unlikely to be
able to account for the microlensing events (Gould, Flynn, \& Bahcall 1998).
Thus a
spheroidal distribution would again correspond to a previously undetected
component.  For such a distribution the total mass is constrained in order
not to conflict with the inner rotation curve of the galaxy, which limits
LMC optical depths $\tau\la 1.2 \times 10^{-7}$.

The extended protodisk supports approximately half of the local
rotation speed, with the remainder coming from the thin disk
and dark (non-MACHO) halo (see e.g. Figure 1).  The dark halo in these
models has a large core radius ($> 7 \rm{kpc}$) and an asymptotic
rotation speed of $\approx 180 \rm{kpc}$. The total mass in the
Galaxy out to 50 kpc is $\approx 4.6 \times 10^{11}M_\odot$.
For a total mass in the white dwarf extended protodisk
of $M_{wd} = 7\times 10^{10}M_\odot$, we find:

\begin{itemize}
\item The optical depth toward the LMC generated by this component is  
$\tau\sim 1 -  1.5 \times 10^{-7}$; 
\item The lens mass estimates for the current MACHO event durations is
$m\sim 0.5 M_\odot$, consistent with white dwarf masses;
\item We expect to see roughly twice as many white dwarfs in the HDF-South
compared to HDF-North, similar to the halo models;
\item Simulations of the
proper motions for white dwarfs in this model
are broadly consistent
with the observations of Ibata et al. (1999). 
\end{itemize}

The main feature of this model, however, is that it has a much lower total
mass in white dwarfs than halo models. As outlined above, it is consistent
with both the MACHO data and the HDF studies for a total mass in white
dwarfs of $M_{wd}\la  7\times 10^{10}M_\odot$.  This is approximately half of
the mass that would be required for a halo distribution of MACHOs which
would produce the same optical depth.  Basically, this reduction can be
understood because most microlensing is due to lenses within about $20$
kpc of the Sun for either configuration.  The extended protodisk has less
mass beyond that distance than a halo.

The smaller mass in white dwarfs today also implies a smaller total
mass in the progenitor population.  For our above example the progenitor
mass $M_{Pstars}\sim 3\times 10^{11} M_\odot$, a crucial factor of 5
less than that for a $20\%$ white dwarf halo, assuming the same IMF in
both cases.

There are several predictions of this model that can eventually allow it
to be distinguished from a standard halo white dwarf population.
First, the LMC optical depth cannot be much greater than about 
$1.5\times 10^{-7}$.
Second, because the lenses are concentrated closer to the plane of the galaxy,
the typical lens-observer distance will be smaller (of order $5$ kpc).  
This in turn implies an increase in the expected number of parallax events
(Gyuk \& Gates 1999).
Finally, the ratio of optical depths toward the Small and Large Magellanic
Clouds is expected to be of order $\tau_{SMC}/\tau_{LMC} \sim 0.8$, in contrast
with $\tau_{SMC}/\tau_{LMC} \sim 1.5$ (Sackett \& Gould 1993) predicted for a 
standard halo. 

The distribution of event durations and a detailed
comparison of the distribution (in direction and magnitude) of the
observed proper motions will also differ for a halo vs. extended protodisk
white dwarf population, but these seem unlikely to be able to
differentiate between models without significantly more data.

\section{Model Implications}
 
Because the total mass in the white dwarf population today is significantly
lower in this model, some of the constraints on a halo white dwarf population
can be evaded, including those which consider the progenitor population
and the ejected metal enriched gas. The total mass in this new
component represents less than about $4\%$ of a total halo mass of
$2\times 10^{12}$.  Since essentially all of the current constraints
on white dwarf halos which limit the halo mass fraction in white dwarfs
do so at only the $10\%$ level, these constraints can be satisfied by
our model. This includes the Type Ia supernovae constraints
which are dependent on the mass in white dwarfs today, and 
cannot be evaded by scenarios which involve somehow hiding the
metal enriched gas produced by the progenitor  stars.

However, there remains much work to be done to more carefully
consider the implications of this new component. 
First, we still require 
an IMF which differs significantly from the
disk IMF.  Assuming a log-normal distribution,
Adams \& Laughlin (1996) and Chabrier et al. (1996) 
used conservative constraints to limit the mass fraction
of the high and low mass 
end of the progenitor IMF for a halo white dwarf population.
While the lower total mass in our progenitor population will
relax the constraints (which are based on
the metallicity of the Galactic disk) somewhat at the high mass end, 
the mass fraction of low mass 
($m< 1 M_{\odot}$) stars is constrained by number counts of
faint low mass stars locally. The extended protodisk has
an increased local density relative to a halo distribution,
but a lower total mass, resulting in a constraint
similar to that for a halo. Thus we expect to require a
fairly sharp low mass drop-off in the progenitor IMF.
The implications of such an IMF, 
including the lower fraction of primordial baryons which is processed 
through this early population, need to be examined in greater depth.

This new component also provides some intriguing hints for cosmology.
When did this component form and how is it related to
galaxy formation scenarios?  Can this early starburst population
help us to trace the baryons in the Universe from their primordial
state to the present, where we find most of the baryons in the intracluster 
medium?   
 
\section{Conclusions}
 
We have argued that the microlensing data toward the LMC, combined with
observations of white dwarf stars in a proper motion study of the HDF
indicate the presence of a new component of the galaxy.  This component
can be generally described as an extended distribution
that extends at least 2 kpc above the galactic plane, but resides well
within the dark matter halo.  It is consistent with all data and observations
of the structure and kinematics
of the galaxy, and significantly alleviates the considerable problems
with a halo population of white dwarf stars that is consistent
with microlensing data.  Much work remains to carefully consider the
implications of such a component, in particular the formation and
evolution of the early population
of progenitor stars (and resulting metal enriched gas)
that produced this component. However, the significantly lower mass
in the progenitor population as compared to that for a halo population of
white dwarfs will allow a reasonable fraction of the baryonic
mass of the Universe to remain in gas that has not been processed through
these very early stars.  Moreover, this component may be a more reasonable
distribution for the remains of an early starburst population, in which
one would expect a more condensed distribution than that of the halo.

\acknowledgments
 
This work was supported in part by the DOE (at Chicago and Fermilab)
and by the NASA (through grant NAG5-2788 at Fermilab).

\end{document}